\documentstyle[prb,aps,epsf,multicol] {revtex}
\begin{document}
\bibliographystyle{prbsty}
\draft
\title{Theoretical analysis of the experiments on the 
       double-spin-chain compound -- KCuCl$_3$}
\author{Tota Nakamura$^1$ and Kiyomi Okamoto$^2$ }
\address{$^1$ Department of Applied Physics, Tohoku University,
         Sendai, Miyagi 980-77, Japan\\
         $^2$ Department of Physics, Tokyo Institute of Technology,
         Oh-Okayama, Meguro, Tokyo 152, Japan
         }
\date{\today}
\maketitle
\begin {abstract}
We have analyzed the experimental susceptibility data of 
KCuCl$_3$ and found that the data are well-explained by the 
double-spin-chain models with strong antiferromagnetic dimerization.
Large quantum Monte Carlo calculations were performed for the 
first time in the spin systems with frustration.
This was made possible by removing
the negative-sign problem with the use of the dimer basis that has
the spin-reversal symmetry.
The numerical data agree with the experimental data 
within 1\% relative errors in the whole temperature region.
We also present a theoretical estimate for the dispersion relation
and compare it with the recent neutron-scattering experiment.
Finally, the magnitude of each interaction bond is predicted.
\end  {abstract}


\begin{multicols}{2}

\narrowtext

The low-dimensional quantum systems
with the excitation energy gap are now attracting much interest
both experimentally and theoretically.\cite{dagotto-r96}
Possibility of the high-T$_{\rm c}$ superconductivity upon doping
carriers to the gapped insulator lies as the background.
The spin-ladder model 
and its realization in a real compound SrCu$_2$O$_3$
may be the most well-known substance for this scenario.
\cite{azuma-htik94}
Of course, 
various other new compounds are also synthesized which
can be explained by the low-dimensional quantum spin Hamiltonian.
\cite{ramirez94,tanaka-tso96,onoda-n96,takatsu-st97%
,hammar-rb97,tennant-nbgrs97}
These experimental achievements
now offer many data left for theoretical investigations.

Tanaka {\it et al}. \cite{tanaka-tso96} 
have measured the magnetic susceptibility of the
KCuCl$_3$, and proposed that it 
can be explained by the double-spin-chain Hamiltonian. 
The susceptibility data show the spin-gap behavior. 
They estimated 
the amount of the excitation gap $\Delta$
by fitting the low-temperature data with its
theoretical expression,
$\chi(T) \propto T^{-1/2}\exp(-\Delta /T)$,
given by Troyer {\it et al}.\cite{troyer-tw94}
The estimated gap, $\Delta/k_{\rm B}\sim 35$K, is
consistent with their recent measurement of the magnetization process
at low temperature, which gives $\Delta/k_{\rm B}\sim 31.1$K.
\cite{shiramura-ttktmg97}

For the thorough understandings of the system, we are necessary to
determine the strength of each interaction bond of the model Hamiltonian
which expresses the subject material.
For frustrated systems, one usually calculates all the eigenvalues of a
given model Hamiltonian on a finite lattice by the numerical diagonalization,
and then compares the obtained 
physical quantities at finite temperatures with the experimental data.
Since the numerical diagonalization technique is restricted to the 
very small sizes with sixteen $S=1/2$-spins or the less,
the results suffer a severe finite size effect.
The quantum Monte Carlo (QMC) method can handle much larger systems,  but
has not been applied to the frustrated systems
because of the negative-sign problem.
Recently, one of us (T. N.) has solved this problem 
in the double-spin-chain system.\cite{tota97}
It makes possible to treat large system sizes even at very low temperatures,
and thus we are able to compare directly 
the numerical results to the experimental 
data without suffering the size effect.

In this Rapid Communication, 
we calculate the susceptibility of the double-spin-chain
model with 162 spins,
and aim at the determination of the interaction bonds that explain KCuCl$_3$.
For this purpose, it is necessary to calculate the dispersion relation 
of the excited state to compare with the neutron-scattering 
experimental results. 
The susceptibility alone is not sufficient 
to discuss the details of the system, since it is the integrated quantity.
We use our analytic expression for the dispersion relation obtained recently
for general double-spin-chain models.\cite{tota-tok97}
Our analysis partly disagrees with the experiments, which may be understood
by possible two-dimensional couplings.

\begin{figure}
    \epsfxsize = 8.5cm
    \epsffile{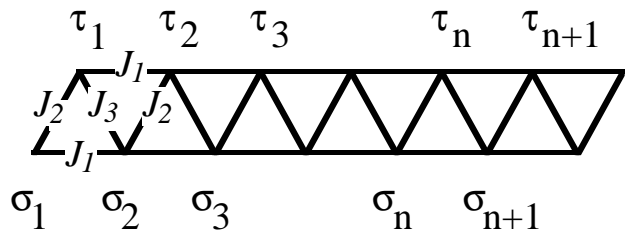}
 \caption {Shape of the general double spin-chain model.
  \label{fig:lattice}
          }
\end  {figure}

We consider the following spin Hamiltonian under the open 
boundary conditions:
\begin{eqnarray}
 {\cal H}=\sum_{n=1}^{N-1} \Big\{
    &J_1&(\mbox{\boldmath $\sigma$}_n \cdot\mbox{\boldmath $\sigma$}_{n+1}
  +     \mbox{\boldmath $\tau  $}_n \cdot\mbox{\boldmath $\tau  $}_{n+1})
  +  J_2  \mbox{\boldmath $\sigma$}_n \cdot\mbox{\boldmath $\tau  $}_{n}
\nonumber \\
  + &J_3& \mbox{\boldmath $\tau  $}_n \cdot\mbox{\boldmath $\sigma$}_{n+1}
  \Big \}
  +  J_2  \mbox{\boldmath $\sigma$}_{N}\cdot\mbox{\boldmath $\tau  $}_{N}.
\label{eq:hamiltonian}
\end  {eqnarray}
Here, $N$ is the linear size of the system, 
and $|\mbox{\boldmath $\sigma$}|=|\mbox{\boldmath $\tau$}|=1/2$.
Figure \ref{fig:lattice} shows the shape of the lattice.

Three cases of the above Hamiltonian,
(i) $J_1=J_3$, (ii) $J_3=0$, and (iii) $J_1=0$,
are especially investigated.
In each case, $J_2$ is set variable ranging from $J_2=-8$ to $J_2=8$.
The first one, $J_1=J_3$, is what we call here the `zig-zag'
or commonly called the railroad-trestle model.
This model has frustration for $J_2 > 0$, and thus has not been analyzed
on its thermodynamic properties yet.
The second one is the ordinary two-leg `ladder' model, and
the third one is the `bond-alternation' model.
By changing the sign of the $J_2$-bond, 
the Hamiltonian can express 
both the dimer-gap system and the Haldane-gap system.
Difference in the origin of the gap affects the structure of the 
excited states and thus the finite temperature behavior of 
various physical quantities.
We also comment on this point at the end of this Rapid Communication.

Before the demonstration of the numerical evidences, we briefly summarize the
simulational technique.
We have done the ordinary world-line QMC simulations, but
the choice of the representation basis is considered different from the 
conventional $s^z$ one.
Two spins, $\mbox{\boldmath $\sigma$}_n$ and 
$\mbox{\boldmath $\tau$}_n$, are coupled and considered as a unit of update.
This dimer unit takes four states associated with the $s^z$ eigenvalues of
each spin, $|\sigma^z, \tau^z\rangle$.
We rearrange these four states so that they have the spin-reversal symmetry:
\begin{eqnarray}
 v_1 &=& (|\uparrow, \uparrow\rangle + |\downarrow, \downarrow\rangle)/
\sqrt{2},\\
 v_2 &=& (|\uparrow, \uparrow\rangle - |\downarrow, \downarrow\rangle)/
\sqrt{2},\\
 v_3 &=& (|\uparrow, \downarrow\rangle + |\downarrow, \uparrow\rangle)/
\sqrt{2},\\
 v_4 &=& (|\uparrow, \downarrow\rangle - |\downarrow, \uparrow\rangle)/
\sqrt{2}.
\end  {eqnarray}
Here, $\uparrow$ and $\downarrow$ denote the $s^z$ eigenstates.
Now,
two $J_1$-bonds and one $J_3$-bond just become a single effective bond
connecting the neighboring dimer units,
and a $J_2$-bond only contributes to the inner energy of a dimer unit.
We can remove the negative-sign problem by using this representation basis.
The Trotter number $m$ are chosen so that 
$\beta (2J_1+J_3)/m$ varies from 0.35 to 0.15 at each temperature,
and the extrapolations to the infinite $m$ are done.
Typical number of the Monte Carlo steps is 500 000 divided into ten parts
to see the statistical deviation.
The first 50 000 steps are discarded for the equilibrium.
The autocorrelation time of the susceptibility is less than an 
order of unity.

We have performed simulations for several values of $J_2$ in each choice of
the $J_1$- and the $J_3$- bonds.
First, we have set $J_1=J_3$ and done simulations thoroughly by changing 
a value of $J_2$ from $-8$  to $8$ in a step of 2.
A rough estimate for the value of $J_2$ is made at this stage.
Then, we have proceeded to the other cases, $J_1=0$ and $J_3=0$.

\begin{figure}
    \epsfxsize = 8.5cm
    \epsffile{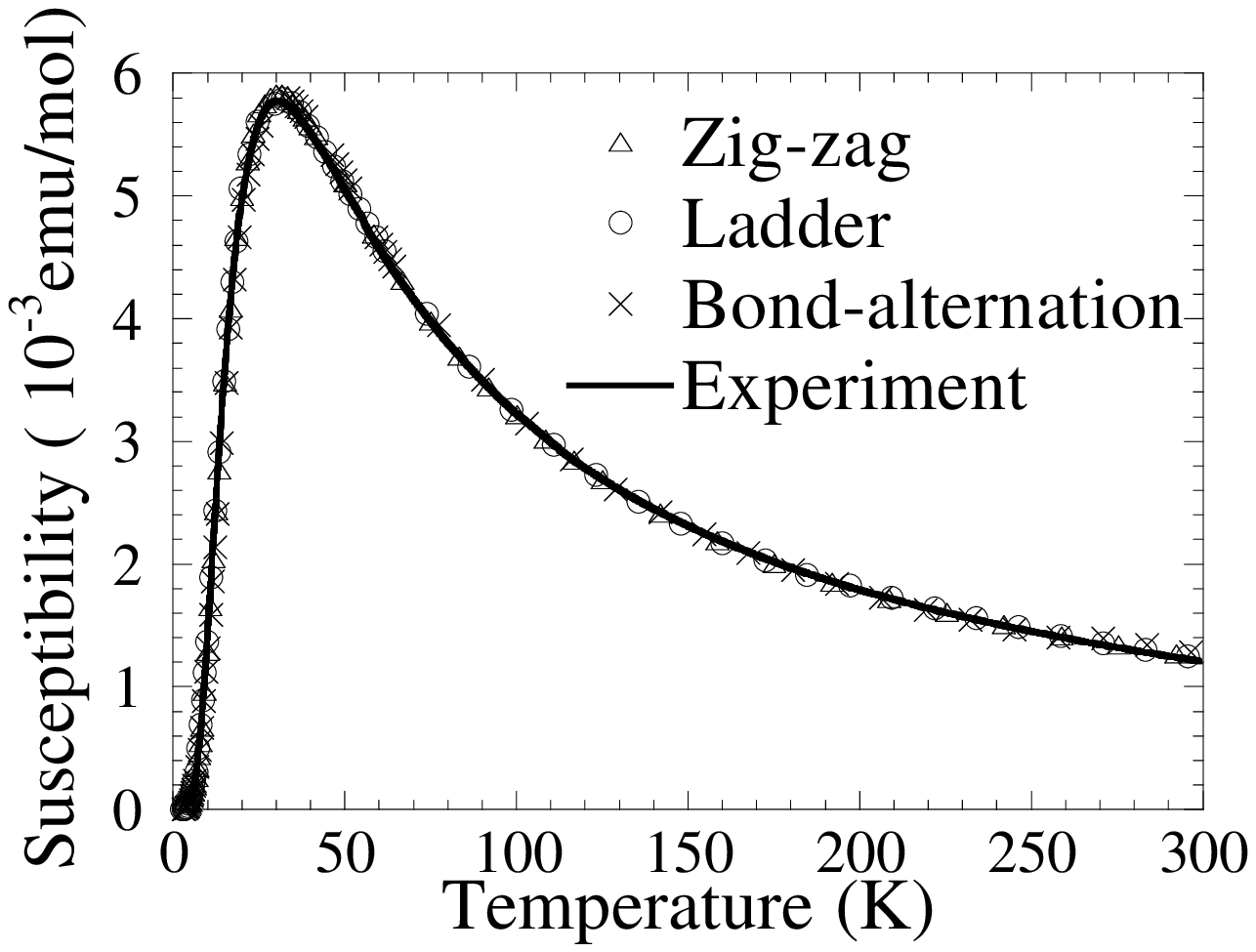}
    \epsfxsize = 8.5cm
    \epsffile{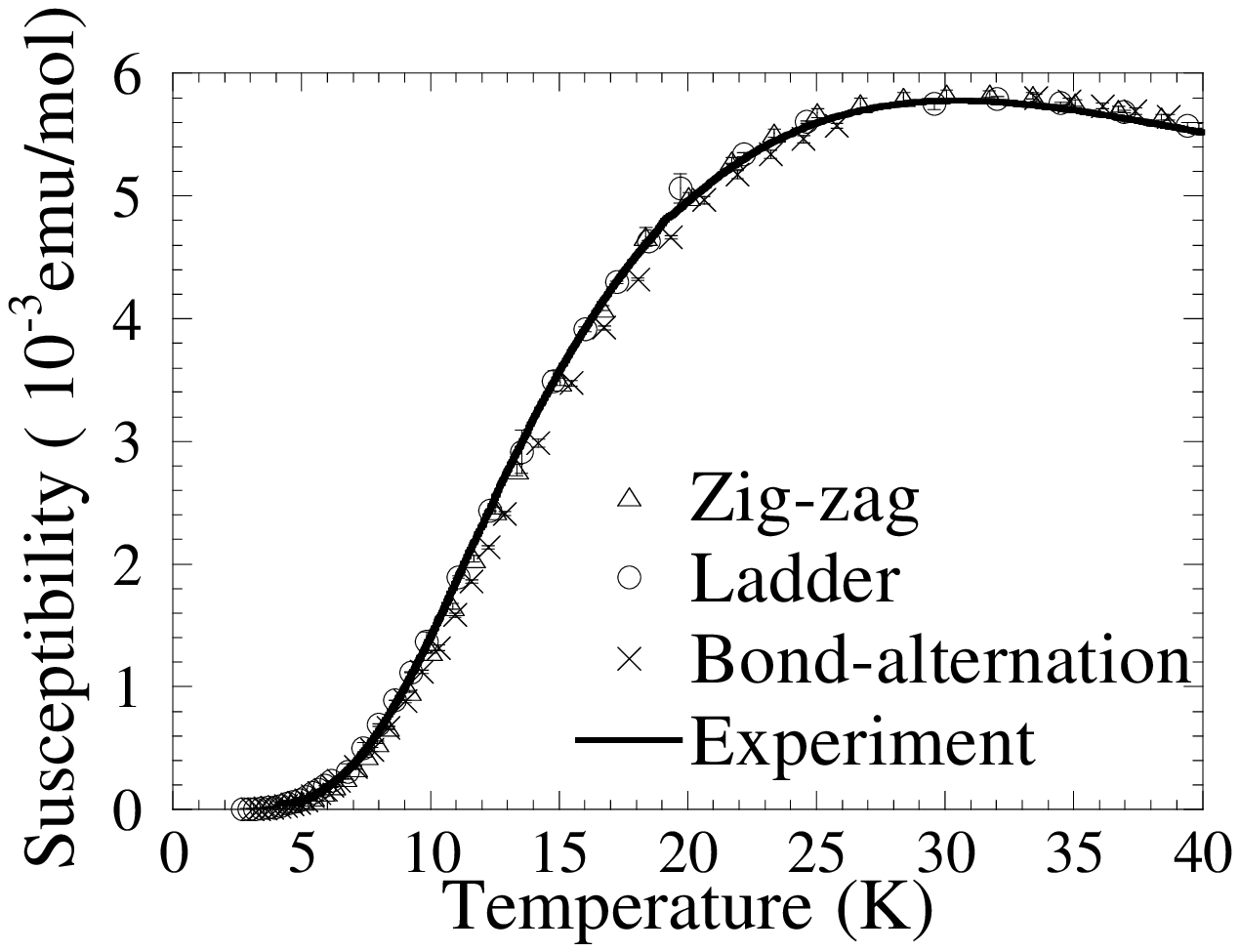}
 \caption {(a) The QMC results of the susceptibility (symbols) compared with
           the experimental data (line); (b) its low-temperature plot.
           The number of the spins is 162 and the open boundary conditions
           are used.
           The strength of each interaction bond is;
           ($\bigtriangleup$)  Zig-zag: $J_1=J_3=8.35$K, $J_2=6J_1$,
                                        $g=2.05$,
           ($\circ$) Ladder: $J_1=12.3$K, $J_3=0$, $J_2=4J_1$,
                                        $g=2.05$, and
           ($\times$) Bond-alternation: 
                             $J_3=12.9$K, $J_1=0$, $J_2=4J_3$, $g=2.03$, 
           respectively.
  \label{fig:sus}
          }
\end  {figure}

Figure \ref{fig:sus}(a) shows the QMC results of the susceptibility
compared with the experimental data,
and (b) is its low-temperature plot.
The experimental data are the minimum susceptibility when 
the magnetic field is applied in the cleavage plane.
The absolute value of the susceptibility depends upon the direction 
of the field, which can be understood by considering the anisotropy 
of the $g$-value.\cite{tanaka-tso96}
Here, the $g$-value is determined by the comparison at high temperatures,
$T > 100$K.
Error bars of the QMC data are almost negligible.
Magnitude of each interaction bond is determined in order that 
the low-temperature data agree quantitatively with the experiment.

Triangles are the best fit to the experiment so far in the case of
$J_1=J_3$, and the
values of the interaction bonds are:
$J_1=J_3=8.35$K, $J_2=6J_1$,
i.e., all the interaction bonds are antiferromagnetic.
The $g$-value $g=2.05$.
Circles are those of the ordinary ladder case,
and the values are: $J_1=12.3$K, $J_3=0$, $J_2=4J_1$, and $g=2.05$.
Crosses are the case of the bond-alternation model:
$J_1=0, J_3=12.9$K, $J_2=4J_1$, and $g=2.03$.
All three data quantitatively agree with the experiment 
at all the temperatures.
Our estimate for the $g$-value is also consistent with the ESR-measurement
giving $g=2.05$.\cite{shiramura-ttktmg97}
Relative errors from the experiment are within $1\%$ in every case,
and are smallest in the `ladder' case.
However,
we are not sure the small deviations among three cases are relevant or not, 
because we have used a very-simplified model spin-Hamiltonian for
a real compound.
Systematic errors from an adoption of this model may be the most 
important one.
Therefore, we cannot determine the strength of each interaction bond 
from the fitting of the susceptibility data alone.
Only a common feature known from this plot is 
that the $J_2$-bond is strongly
dimerized antiferromagnetically,
and thus the origin of the gap is the dimer gap.

We try to determine the interaction bonds by the dispersion relation 
of the excitation energy now.
This value is quite sensible to the details of the model.
Recently, Kato {\it et al}. \cite{kato-ttsnk97}
applied the neutron-scattering analysis 
on this compound, and presented the dispersion relation.
We compare our analytic dispersion relation with this experiment.

In our previous paper,\cite{tota-tok97}
we have deduced an expression giving the dispersion of the 
general double-spin-chain systems as a function of the interaction bonds.
There, we employed a single domain-wall variation after the non-local unitary 
transformation \cite{kennedy-t92,takada-k91} is applied.
This approximation is quite excellent near $k=\pi$, checked with the 
numerical diagonalization results of finite systems, and with the 
exactly-known results.
On the other hand,
it becomes overestimated as the wave number $k$ approaches zero.

\begin{figure}
    \epsfxsize = 8.5cm
    \epsffile{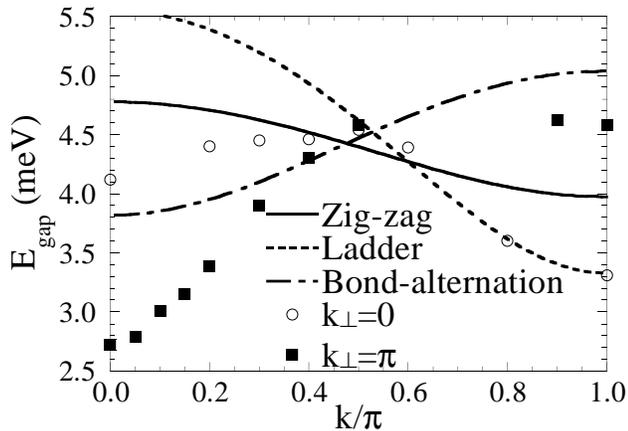}
 \caption {The dispersion relation of the excited states of the system.
           Zig-zag: $J_1=J_3=8.35$K and $J_2=6J_1$,
           Ladder: $J_1=12.3$K, $J_3=0$, and $J_2=4J_1$, and
           Bond-alternation: $J_3=12.9$K, $J_1=0$, and $J_2=4J_3$, 
           respectively.
           The neutron-scattering experimental data are plotted by
           circles for the constant wave number perpendicular to the 
           chain $k_{\perp}=0$, and by solid squares for 
           $k_{\perp}=\pi$.
  \label{fig:disp}
          }
\end  {figure}

By using the values of the interaction bonds obtained from the susceptibility 
fitting, we give the dispersion of the excitation energy
in Fig. \ref{fig:disp}.
The first excitation is at $k=\pi$ in the `zig-zag' and the `ladder' models,
and is at $k=0$ in the `bond-alternation' model.
Within our analysis,
the first excitation is always at $k=\pi$ for $J_1 > \min(J_2, J_3)/2$, 
and is otherwise at $k=0$, where
$\min(J_2, J_3)$ stands for the minimum value of $J_2$ and $J_3$.

The dispersion of the `ladder' case is consistent 
with the experimental dispersion parallel to the double chain 
when the constant wave number perpendicular to the chain $k_{\perp}$
is equal to zero.
Our estimate of the gap is $\Delta / k_{\rm B} \sim 38$K, which is 
a little larger than the results of the magnetization process experiment.
\cite{shiramura-ttktmg97}
So the `ladder' model may be the most favorable candidate for explaining the
KCuCl$_3$ at the present stage.
However, it should be noticed that the experimental dispersion is 
dependent upon $k_{\perp}$.
If a double-chain is isolated from each other, there should not be
the $k_{\perp}$ dependences.
Therefore, it might be an evidence of two-dimensional interaction couplings.
This point is left for the future study.

Tanaka {\it et al}. \cite{tanaka-tso96} fitted their experimental data
of the susceptibility to the theoretical expression
$\chi(T) = (1/2\sqrt{\pi a T}) \exp(-\Delta/T)$,
given by Troyer {\it et al}. \cite{troyer-tw94} supposing the form 
$\epsilon_k = \Delta + a|k-\pi|^2$ for the magnon dispersion.
This expression is valid when $T \ll \Delta, D$, where $D$ is the band width
of the magnon dispersion.
If we take the effect of the band width into account, we obtain
\begin{equation}
  \chi(T)
  = {{\rm e}^{-\Delta/T} \over 2\pi \sqrt{aT}}
    \gamma \left( {1 \over 2}, {D \over T} \right),
\end{equation}
where $\gamma(x,p)$ is the incomplete gamma function defined by
\begin{equation}
  \gamma(x,p) 
  \equiv \int_0^p t^{x-1} {\rm e}^{-t}\,dt .
\end{equation}
Since $D \ll \Delta$ from Fig. \ref{fig:disp}, $\chi(T)$ behaves as
\begin{equation}
  \chi(T) \sim
  \cases{
      \displaystyle{{\rm e}^{-\Delta/T} \over 2\sqrt{\pi a T}}
      \left( 1 -\sqrt{T \over \pi D}{\rm e}^{-D/T} \right) 
      &($T \ll D\ll \Delta$), \cr
      \displaystyle{1 \over \pi}\sqrt{D\ \over a}{{\rm e}^{-\Delta/T} \over T}
      \left( 1 - {3D \over T} \right)
      &($D \ll T \ll \Delta$). \cr
  }
\end{equation}
Thus, the expression of Troyer overestimates
the susceptibility. 
In the fitting by Tanaka,
the theoretical curve of Troyer severely deviates from
the experimental curve for $T > 20 {\rm K}$,
which is well-explained by $D = 2.2 \,{\rm meV} = 25 \,{\rm K}$
of the `ladder' model.

\begin{figure}
    \epsfxsize = 8.5cm
    \epsffile{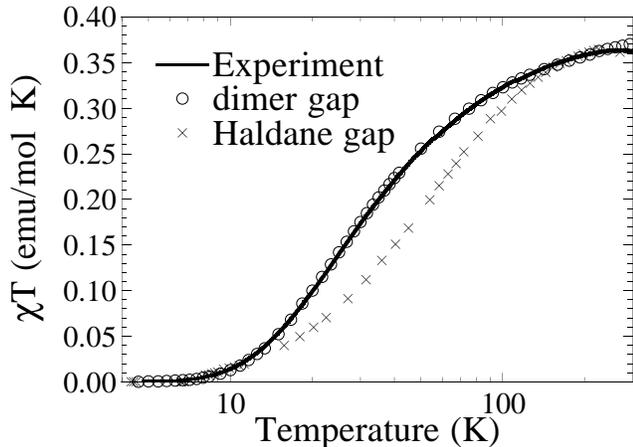}
 \caption {Product of the susceptibility and the temperature, $\chi T$, is
           plotted against the temperature.
           The QMC results of the dimer-gap system ($\circ$) 
           and the Haldane-gap system ($\times$)
           are compared with the experimental data (line).
           The strength of each interaction bond is,
           ($\circ$) : $J_1=J_3=8.35$K, $J_2=6J_1$, $g=2.05$, and
           ($\times$): $J_1=J_3=45$K, $J_2=-4J_1$, $g=2.00$, respectively.
  \label{fig:dimerHaldane}
          }
\end  {figure}

In the last part of this Rapid Communication,
we mention the difference of the susceptibility in the dimer-gap system
and that in the Haldane-gap system.
Only the `zig-zag' case, $J_1=J_3$, is demonstrated as an example
in Fig. \ref{fig:dimerHaldane}.
Product of the susceptibility  and the temperature, $\chi T$, is plotted
against the temperature, since this value is independent from the 
scale of the interaction bond.
We use the logarithmic scale for the temperature axis, so that 
the rescaling of the temperature by changing the values of 
interaction bonds just causes a parallel shift along the 
temperature axis.
Circles are the data of the dimer-gap system:
$J_1=J_3=8.35$K, $J_2=6J_1$, $g=2.05$, and
crosses are those of the Haldane-gap system:
$J_1=J_3=45$K, $J_2=-4J_1$, $g=2.00$.
These interactions are determined by fitting the low-temperature data,
while the $g$-value is estimated at high temperatures.
As seen from this figure, 
function form of $\chi T$ in the Haldane-gap system is obviously different 
from the experiment,
so we cannot fit them by any parallel shift.

Let us consider the dimer limit, $J_1 = J_3 =0$, $J_2>0$,
where the ground state is an array of independent 
singlet dimers and thus is non-degenerate.
When the interactions between singlet dimers ($J_1$ and $J_3$) are introduced,
the ground state is somewhat modified, but still has the nature of 
the singlet dimer.
On the other hand, the ground state in the Haldane limit,
$J_1 = J_3 =0$, $J_2<0$, is an array of independent triplet dimers and is
highly degenerate.
When $J_1$ and $J_3$ are switched on, the degeneracy is lifted up resulting
in the unique ground state.
The density of states of low-lying excited states are larger
than the singlet dimer case, reflecting this high degeneracy
of the ground state at $J_1 = J_3 =0$.
Therefore, the susceptibility peak 
in the Haldane system becomes lower and broader than in the dimer system.
We may determine the origin of the gap from the full width at half maximum
of the susceptibility, or the function form of $\chi T$.

In summary,
we have analyzed the susceptibility 
and the neutron-scattering measurements on the KCuCl$_3$
from the theoretical point of view.
Large-scale quantum Monte Carlo simulations clarified this compound is 
well-explained by the double-spin-chain Hamiltonian with strong 
{\it antiferromagnetic} dimerization.
Since the susceptibility alone is not sufficient to determine all the 
interaction-bond strengths, we also calculated the dispersion relation of 
the excited state to compare with the recent neutron-scattering experiment.
As far as we restrict our theoretical analysis to the 
isolated double-spin-chain Hamiltonian,
the best one to explain both measurements at present is the ladder model,
which is defined by $J_1=12.3$K, $J_2=4J_1$, $J_3=0$, and $g=2.05$.
This $g$-value agrees with the ESR-measurement.\cite{shiramura-ttktmg97}
However, the experimental dispersion relation has strong dependence 
on the constant wave number perpendicular to the chain, $k_{\perp}$,
which cannot be explained by our theoretical analysis on the isolated
double-spin-chain model.
Thus, the system may have two-dimensional interactions that are not
relevant at low temperatures.

The quantum Monte Carlo simulation has become the most realistic method
to investigate the various quasi-one-dimensional compounds that can be
expressed by the general double-spin-chain Hamiltonian
even if the system has frustration.
Present calculations can be extended to analyze other experiments easily. 
For example, we comment on 
that of Cu$_2$(1, 4-Diazacycloheptane)$_2$Cl$_4$ recently done by
Hammar {\it et al}. \cite{hammar-rb97}
Their susceptibility data are
consistent with our calculation down to 
the lowest temperature that could not be obtained by the 
numerical diagonalization.
The choice of the interactions is $J_1=J_3=2.26$K, $J_2=6J_1$, and
$g=2.04$.

Authors 
would like to thank H. Tanaka, T. Kato, and D. H. Reich
for valuable discussions and for
sending us the experimental data.
They also acknowledge thanks to H. Nishimori for his 
diagonalization package Titpack Ver. 2,
and to N. Ito and Y. Kanada for their random-number generator RNDTIK.
A part of the computations were carried out
on Facom VPP500 at the ISSP, University of Tokyo.

\end{multicols}


\end{document}